\documentclass[aps, prx, longbibliography,twocolumn,showpacs,superscriptaddress]{revtex4-1}  
\usepackage{graphicx}  
\usepackage{dcolumn}   
\usepackage{bm}        
\usepackage{amssymb}   
\usepackage{amsmath}   
\usepackage{hyperref}
\usepackage{bbm}
\usepackage[caption=false]{subfig}
\usepackage{graphicx}
\usepackage{sidecap}

\usepackage{xcolor}

\usepackage{appendix}
\def\01{\{0,1\}}

\newcommand{\microspace}{\mspace{0.5mu}}

\newcommand{\floor}[1]{\left\lfloor #1 \right\rfloor}

\def\({\left(}
\def\){\right)}

\def\<{\langle}
\def\>{\rangle}
\def \lket {\left|}
\def \rket {\right\rangle}
\def \lbra {\left\langle}
\def \rbra {\right|}
\newcommand{\ket}[1]{\lket\microspace #1 \microspace\rket}
\newcommand{\bra}[1]{\lbra\microspace #1 \microspace\rbra}

\newcommand{\braket}[2]{\langle #1 \microspace | \microspace#2 \rangle}

\begin{document}
\title{Polynomial time algorithms for estimating spectra of adiabatic Hamiltonians}
\author{Jacob~Bringewatt}
\affiliation{Department of Physics, University of Maryland, College Park, Maryland 20742, USA}
\affiliation{Joint Center for Quantum Information and Computer Science, College Park, Maryland 20742, USA}
\affiliation{Joint Quantum Institute, College Park, Maryland 20742, USA}
\author{William~Dorland}
\affiliation{Department of Physics, University of Maryland, College Park, Maryland 20742, USA}
\author{Stephen~P.~Jordan}
\affiliation{Microsoft Quantum, Redmond, Washington 98052 USA}
\affiliation{University of Maryland Institute for Advanced Computer Studies, College Park, Maryland 20742, USA}
\date{\today}

\begin{abstract}
Much research regarding quantum adiabatic optimization has focused on stoquastic Hamiltonians with Hamming symmetric potentials, such as the well studied ``spike" example. Due to the large amount of symmetry in these potentials such problems are readily open to analysis both analytically and computationally. However, more realistic potentials do not have such a high degree of symmetry and may have many local minima. Here we present a somewhat more realistic class of problems consisting of many individually Hamming symmetric potential wells. For two or three such wells we demonstrate that such a problem can be solved exactly in time polynomial in the number of qubits and wells. For greater than three wells, we present a tight-binding approach with which to efficiently analyze the performance of such Hamiltonians in an adiabatic computation. We provide several basic examples  designed to highlight the usefulness of this toy model and to give insight into using the tight-binding approach to examining it, including: (1) adiabatic unstructured search with a transverse field driver and a prior guess to the marked item and (2) a scheme for adiabatically simulating the ground states of small collections of strongly interacting spins, with an explicit demonstration for an Ising model Hamiltonian.   
\end{abstract}
\maketitle
\begin{section}{Introduction}
Since their introduction in \cite{Bravyi}, so-called \emph{stoquastic} Hamiltonians, those with real non-positive off-diagonal matrix elements, have been a major point of focus for research regarding adiabatic quantum computation (AQC)\cite{Farhi}. Stoquastic Hamiltonians have real, non-negative ground states, and therefore, a question of particular interest is whether AQC with stoquastic Hamiltonians is capable of exponential speedup over classical computation. We note that the computational cost for an AQC problem is determined via adiabatic theorems which upper bound the runtime as the inverse of the eigenvalue gap between the lowest two eigenvalues squared~\cite{jansen2007bounds, elgart2012note}. As a complexity theory question, the computational power of stoquastic AQC is still unknown, but for specific classical algorithms such as path integral \cite{Hastings} or diffusion Monte Carlo (MC) \cite{jarret2016adiabatic, Bringewatt}, examples have been presented where exponential speedup \textit{over the specific classical algorithm} is indeed still possible with stoquastic Hamiltonians. 

However such finely tuned examples raise new questions as to whether such obstructions to classical simulation are typical of more general stoquastic Hamiltonians. The diffusion MC examples and others, such as the well studied ``spike" example \cite{farhi2002quantum,brady2016spectral, R04, CD14, MAL16, Bringewatt} take advantage of heavy symmetry with a potential that is a function of Hamming weight to allow for efficient analytic and computational analysis. It has been shown that such problems are always efficiently classically simulatable by path integral quantum Monte Carlo (QMC) \cite{jiang2017scaling}, which suggests somewhat more complicated models would be helpful for addressing these questions.

Here we consider such a model designed to be both efficiently analyzable and somewhat more realistic than purely Hamming symmetric problems. In particular, we expect realistic cost functions to have many local minima; therefore we consider a collection of $K=\mathrm{poly}(n)$ individually Hamming symmetric wells. While the full Hamiltonian in such a model is no longer Hamming symmetric, enough symmetry remains to allow for a similar reduction to an effective polynomial sized subspace. We show that this reduction can always be exact for $K=2, 3$.

For larger $K$ we introduce an approximate tight-binding scheme for analyzing the model. The reduction this model affords is represented diagramatically in Fig. \ref{fig:tbreductionrep} for three Hamming symmetric potential balls on 10 qubits. In addition to being computationally efficient, this tight-binding model makes the effects of tunneling readily apparent; in the tight-binding model, the minimum eigenvalue gap is in many cases dictated by the matrix element between ground states of neighboring potential wells, which in turn is dominated by the ``tunneling" part of the wavefunctions, i.e. the amplitudes on bit strings for which the potential energy is greater than the eigenenergy of the state. As interference is not manifest in the ground state of stoquastic Hamiltonians, it is expected that if AQC with such Hamiltonians is to provide advantages over classical computation these advantages should lie in the power afforded by tunneling effects between local minima of the cost function; therefore even if the model presented here also proves to be efficiently simulatable classically, it still provides a useful new tool set for addressing and understanding the performance of AQC with more realistic stoquastic Hamiltonians.

In this paper we present this model and our exact and approximate algorithms for analyzing it, along with a collection of examples designed to highlight their strengths and limitations.

\begin{figure}
    \centering
    \includegraphics[width=9cm,height=9cm,keepaspectratio]{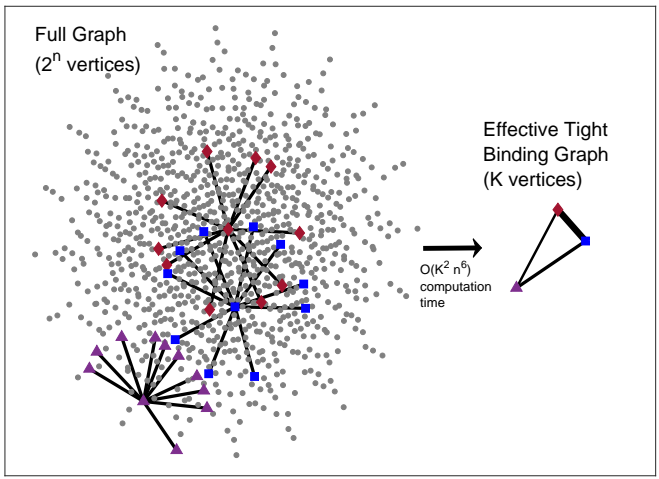}
    \caption{The key feature of our model and the tight-binding approach to analyzing it is the ability to exponentially reduce the size of the problem to an effective Hamiltonian that explicitly considers tunneling between a collection of Hamming symmetric wells. Here we diagramatically demonstrate this reduction for a collection of 3 Hamming symmetric wells of width 1 on 10 qubits. The original graph representing our Hamiltonian has a hypercube geometry, but the edges not within wells are eliminated for visual clarity.}
    \label{fig:tbreductionrep}
\end{figure}
\end{section}
\begin{section}{Tight-binding approximation}
Hamiltonians with $-\sum_j X_j$ driving terms and Hamming-symmetric potentials comprise a common class of Hamiltonians considered in AQC \cite{farhi2002quantum,brady2016spectral, R04, CD14, MAL16, Bringewatt, jarret2016adiabatic}, where Hamming weight is defined as the number of ones in a bit string corresponding to a basis state of the Hilbert space. Such Hamiltonians are often used due to the ability to block diagonalize the Hamiltonian into smaller subspaces: if we think of our qubits as spin-1/2 particles each block corresponds to a possible total angular momentum $j$ and we can parameterize within each block by the z-projection of the angular momentum $m$ and a parameter $\gamma$ labeling the degeneracies of the $j$, $m$ representations. 

Here we introduce a reparameterization $\sigma=n/2-j\in[0, n/2]$, $w=n/2+m\in[\sigma, n-\sigma]$ which indexes the permutation symmetric subspace as the $\sigma=0$ subspace and defines $m$ in terms of Hamming weight. The ground state and thus the relevant spectral gap of such Hamiltonians is guaranteed by the Perron-Frobenius theorem to lie in the exponentially-reduced permutation symmetric $\sigma=0, \gamma=0$ subspace. 

As we use the Perron-Frobenius theorem several times throughout this paper, we restate it here for convenience: Let $A$ be a matrix with all real and non-negative entries. Then A has a unique leading eigenvalue with corresponding eigenvector with all elements strictly positive. 

In the context of stoquastic AQC, we have a Hamiltonian with all non-positive matrix elements so $\exp(-\beta H)\simeq 1-\beta H$ is nonnegative for sufficiently small $\beta>0$ and the Perron-Frobenius theorem applies which guarantees a nondegenerate, real, nonnegative ground state. Furthermore, as the ground state is nondegenerate, it is guaranteed to transform in a one dimensional representation of any symmetry group of the Hamiltonian. In particular, Hamming-symmetric Hamiltonians have the symmetry group $S_n$, which has only two one-dimensional representations: the trivial representation and the sign representation. By the positivity of the amplitudes of the ground state, it cannot transform according to the sign representation and therefore is invariant under all permutations of the qubits.

Therefore, analyzing the Hamming symmetric problem in this subspace which has a basis parameterized by the Hamming weight $w$ enables efficient computation of the spectral gap. A detailed review of this reduction is presented in App.~\ref{app:HammingSymmetric}. 

Here, we consider a generalization of the standard problem. Instead of a fully Hamming-symmetric Hamiltonian, we specify $K$ bit strings, each with a Hamming-symmetric potential ``well" about it. This Hamiltonian is of the form
\begin{equation}\label{eqn:generalHamiltonian}
    H(s)=-\frac{a(s)}{n}\sum_j X_j + \sum_{k=0}^{K-1} b_k(s)V_k\Big (\sum_jX^{\bar k}\bar Z_jX^{\bar k}\Big )
\end{equation}
where $X^{\bar k}=\bigotimes_{i=1}^{n}X^{\bar k_i}$ is a bit shift operator shifting a bit string from $\bar k\in\{ 0,1\}^n$ to the all zeros bit string and $\bar Z=(I-Z)/2=\ket 1\bra 1$ is the Hamming weight operator. The remaining (standard) notation is introduced and defined in App.~\ref{app:HammingSymmetric}. Note, for example, that the Grover Hamiltonian with $K$ marked items is a special (simple) case of this Hamiltonian, with $V_k=-1$ for $\sum_j X^{\bar k}\bar Z_j X^{\bar k}\ket{x}=0$ and 0 otherwise. 

Despite the loss of full Hamming symmetry, a similar reduction of Hamiltonians of this form exists, making it possible to calculate the spectral gap of the full problem efficiently. Relabeling symmetries that make this calculation exact for $K=2,3$ are described in App.~(\ref{app:few}). Here we simply indicate a key notation from these exact results for the case of $K=2$: in analogy with the fully Hamming symmetric case we now label our subspaces by a coordinate pair $(\sigma_1, \sigma_2)$. Basis states are further parameterized by two integers $h_1\in [0,n_1]$ and $h_2\in [0,n-n_1]$ where $n_1$ is the Hamming distance between the two wells and a pair $(\gamma_1, \gamma_2)$ labeling degeneracies of representation. The ground state of the Hamiltonian is guaranteed to be in the $(\sigma_1, \sigma_2)=(0,0)$ subspace. The details are left to the appendix. For $K>3$, we introduce a tight-binding approximation, to which we now turn. 

We first consider each of the $K$ wells individually. The eigenstates for each well can be directly and efficiently calculated, as long as one ignores the existence of the other wells. We denote the ground state of the $k^{th}$ isolated well by $\ket{\psi^{(0)}_k}$, and the set of such ground states by $T^{(0)}\equiv \{\psi^{(0)}_k\}$. Similarly, we denote $T^{(1)}\equiv T^{(0)}\cup \{\psi^{(1)}_k\}$ as the set of ground states and (possibly degenerate) first excited states of the individual wells. 

Our zeroth (first) order tight-binding model ansatz consists of the assumption that the ground state $\ket{\phi^{(0)}}$ and first excited state $\ket{\phi^{(1)}}$ of the full Hamiltonian exist in the span of the elements of $T^{(0)}$ ($T^{(1)}$). Therefore starting with the eigenvalue equation $H\ket{\phi}=E\ket{\phi}$ we can insert the tight-binding ansatz $\ket{\phi}=\sum_j c_j \ket{\psi_j}$ for some coefficients $c_j$ to give for the lowest two energy states
\begin{equation}
    H\sum_{j=0}^{K-1} c_j\ket{\psi_j}=E\sum_{j=0}^{K-1} c_j\ket{\psi_j}.
\end{equation}
Then multiplying through by the complete set $T^{(0)}$ (or $T^{(1)}$) of basis states gives the generalized eigenproblem
\begin{equation}
    \sum_{i,j} c_j H_{ij}^{\mathrm{(TB)}}\ket{\psi_i}=E\sum_{i,j} c_jS_{ij}\ket{\psi_i}
    \label{eqn:tbformula}
\end{equation}
where $H_{ij}^{\mathrm{(TB)}}=\bra{\psi_i}H\ket{\psi_j}$ and $S_{ij}=\braket{\psi_i}{\psi_j}$. Solving this generalized eigenproblem gives a variational solution for the lowest two energy states. 

To calculate the elements $H^{\mathrm{(TB)}}_{ij}$ and $S_{ij}$ we use the exponentially reduced subspaces corresponding to the pair of wells $i,j$ (as described in detail in App. \ref{app:few}) to calculate $\ket{\psi_i}$ and $\ket{\psi_j}$. Once we have the basis states $\ket{\psi_i}$ and $\ket{\psi_j}$, calculating the overlap $S_{ij}$ is then self explanatory. To calculate $H^{\mathrm{(TB)}}_{ij}=\bra{\psi_i}H\ket{\psi_j}$ in this subspace we can exactly write the driver part of the Hamiltonian and the diagonal term corresponding to the wells $i,j$ and then add a correction term to exactly include the effects of the other wells in this matrix element. In particular we can write

\begin{align}
\label{eqn:htb}
    &H^{(TB)}_{ij}= \\ 
    &\bra{\psi_i}H_d + \sum_{h_1, h_2}[ V_i(h_1, h_2) + V_j(h_1, h_2) + V_c(h_1,h_2)]\ket{\psi_j} \nonumber
\end{align}
where $H_d$ is the driver part of the Hamiltonian in the appropriate basis $V_i$ and $V_j$ are the diagonal potential terms corresponding to the $i^{\mathrm{th}}$ and $j^{\mathrm{th}}$ wells and

\begin{equation}
\mathrm{diag}(V_c)=\frac{\sum_{k\neq i,j}\sum_{r_k=0}^{n} N(h_1, h_2,n_1,R_{ik},R_{jk},r_k)V_k(r_k)}{\sqrt{\binom{n_1}{h_1}\binom{n-n_2}{h_2}}}
\end{equation}
where $n_1$, $R_{ik}$ and $R_{jk}$ are the Hamming distance between the $i^{th}$ and $j^{th}$ wells, the $i^{th}$ and $k^{th}$ wells and the $j^{th}$ and $k^{th}$ wells, respectively, $r_k$ is the distance from the $k^{th}$ well and $V_k(r_k)$ is the potential due to the $k^{th}$ well at distance $r_k$. The function $N(h_1, h_2,n_1,R_{ik},R_{jk}, r_k)$ gives the number of points of intersection between Hamming spheres of radius $r_i=h_1+h_2$ and $r_j=(n_1-h_1)+h_2$ centered on the $i^{th}$ and $j^{th}$ wells respectively and the Hamming sphere of radius $r_k$ centered on the $k^{th}$ well. For details on how to calculate $N$ see App.~\ref{app:tbbasis}.

Note that calculating the matrix elements $H^{(TB)}_{ij}$ and $S_{ij}$ is only efficient if we can calculate them only by considering a constant (or polynomial) set of reduced subspaces (fixed or bounded $(\sigma_1^{(ij)}, \sigma_2^{(ij)})$) of the $ij^{\mathrm{th}}$ basis. By the Perron-Frobenius theorem and symmetry, the ground state of any well is guaranteed to be in the $(\sigma_1^{(ij)},\sigma_2^{(ij)})=(0,0)$ subspace. So if we just consider zeroth order tight-binding we must only compute individual ground states in this subspace. If we want to include first excited states as in first order tight-binding, however, we must consider the possibility that those states exist in $(\sigma_1^{(ij)}, \sigma_2^{(ij)}) > (0,0)$ subspaces. 

In App. \ref{app:subspacepf} we prove that the first excited state for a given well is guaranteed to exist in one of the (0,0), (1,0), or (0,1) subspaces and thus limits us to a constant set of polynomially-sized subspaces we must diagonalize for first order tight-binding. Here we give a sketch of the proof and the motivating ideas. To simplify things we note that a single well Hamiltonian can also be written in terms of the standard Hamming symmetric subspaces labeled by $\sigma$ and $\sigma_1^{(ij)}+\sigma_2^{(ij)}\leq\sigma$. That is we can show this result by demonstrating that the first excited state of well belongs in either the $\sigma=0$ or $\sigma=1$ subspace.

Start by considering just the driving term of the Hamiltonian $H_d \propto - \sum_j X_j$. The ground state of $H_d$ is therefore proportional to $\ket{+}^{\otimes n}$. The first excited state is $n$ fold degenerate where one of the $n$ bits is flipped to a $\ket{-}$ state. An equal superposition of these states is a permutation symmetric $(\sigma=0)$ eigenstate, leaving $n-1$ states in the $\sigma= 1$ subspace, each labeled by a different $\gamma$ in our $\ket{w\sigma\gamma}$ basis. We note that while within the $\sigma=0$ subspace this first excited eigenstate is the second lowest eigenvalue, in the $\sigma=1$ subspaces for fixed $\gamma=\gamma'$ each of these eigenstates corresponds to the smallest unique eigenvalue \emph{within these subspaces}. Additionally, as each of these fixed $\gamma=\gamma'$, $\sigma=1$ subspaces is itself a stoquastic matrix, by the Perron-Frobenius theorem each of these candidate first excited states is real and non-negative in its respective subspace. 

If we add a Hamming symmetric well, the degeneracy in the first excited state is broken between the $\sigma=0$ state and the $\sigma=1$ states. Which of these is energetically favored depends on the potential and the relative strength of the driving and potential terms, but the $\sigma > 1$ eigenstates can never have lower energy than these states even following the breaking of the degeneracy. To see this we consider an eigenstate $\ket\psi=\sum_{w\sigma\gamma} \alpha(w,\sigma,\gamma)\ket{w\sigma\gamma}$ with corresponding energy (independent of $w$ for $\alpha(w, \sigma, \gamma)\neq 0$) 
\begin{equation}
    E(s)=-\frac{1-s}{n}\big (r^+C^+ + r^- C^- \big )+sV(w)
\end{equation}
where $r^{\pm}=\frac{\alpha(w\pm 1, \sigma, \gamma)}{\alpha(w, \sigma, \gamma)}$ for all $w, \sigma, \gamma$ and $C^{\pm}$ are standard spin-1/2 raising and lowering coefficients (and functions of $w$ and $\sigma$). Note that the potential term is independent of $\sigma$ and thus does not affect which subspace is energetically favored. However, for a subspace $\sigma$ and a subspace $\sigma'>\sigma$ at fixed $w$,  $C^{\pm}(\sigma'>\sigma)<C^{\pm}(\sigma)$ so if $r^{\pm}$ were independent of $\sigma$ then the $\sigma=0$ subspace would always be favored. Consider the energy difference between the candidate first excited states

\begin{align}
    \Delta E(s)&=E_{\sigma'}(s)-E_{\sigma}(s)\nonumber\\
    &=\frac{1-s}{n}[r^+_\sigma C^+_\sigma+r^-_\sigma C^-_\sigma-r^+_{\sigma'} C^+_{\sigma'}-r^-_{\sigma'} C^-_{\sigma'}].
    \label{deltaE}
\end{align}
This equation is independent of $w$ so we take $w=\sigma'$ so that $C^-_{\sigma'}(w=\sigma')=0$ to  eliminate one term. For $\sigma >0$, $r^\pm$ must be nonnegative (by the Perron-Frobenius theorem) so $\Delta E(s)$ is nonnegative unless $r^+_{\sigma'}$ is large relative to $r^\pm_{\sigma}$. By analyzing the eigenvector equation in both subspaces, however, and using the fact that $C_{\sigma'}^+<C_{\sigma}$ we obtain

\begin{equation}
    \frac{E_{\sigma'}-sV }{E_{\sigma}-sV }>\frac{r^+_{\sigma'}}{r^+_\sigma}.
\end{equation}
Both sides are positive definite for $\sigma,\sigma'>0$. And as $sV$ is the same in both subspaces, if $E_{\sigma'}<E_\sigma$ then $r^+_{\sigma'}<r^+_\sigma$ but this contradicts that $r^+_{\sigma}<r^+_{\sigma'}$. Therefore the first excited state must always exist either in the $\sigma=0$ or $\sigma=1$ subspaces. 

Additionally, we can see how the $\sigma=1$ subspace may be energetically favored over the $\sigma=0$ subspace: the $\sigma=0$ subspace does not have the positive definite restriction on $r^{\pm}$, so therefore if there is a sufficiently rapid sign change in the first excited state wavefunction in the $\sigma =0$ subspace as we may see in a bound state of a well, then the $\sigma=1$ subspace is energetically favored. 

Thus, to perform first order tight-binding we must simply check the $\sigma=0$ and $\sigma=1$ subspaces for the first excited state for each well. If the $\sigma=1$ states are energetically favored we use all $n-1$ degenerate first excited states for that well in the tight-binding calculation. Therefore the tight-binding matrices for first order tight-binding can be up to $nK\times nK$ in size.

Before discussing numerical details of the practical implementation of this tight-binding framework, note that this framework explicitly considers the tunneling between potential wells. In particular non-zero off diagonal elements of the matrices $H^{(TB)}$ and $S$ are due to evanescent portions of the bound state wavefunctions that extend beyond their respective wells. When the gap is small the ground state wave function can vary rapidly with $s$ as depicted in Figure \ref{fig:wf}. To understand how this relates to the standard conception of tunneling note that we can make the driver term correspond to our intuitive conception of a kinetic energy term by pulling out an $s$ dependent diagonal term $\propto I$ from the second term in \ref{eqn:generalHamiltonian} to obtain the standard normalized graph Laplacian for this system $H_k=(1-s)[I-\frac{1}{n}\sum_j X_j]$ which can be considered a kinetic energy term \cite{jarret2016adiabatic}. The remaining diagonal potential term is the corresponding potential energy.  

\begin{figure}[h]
    \centering
    \subfloat[]{\includegraphics[width=0.5\columnwidth]{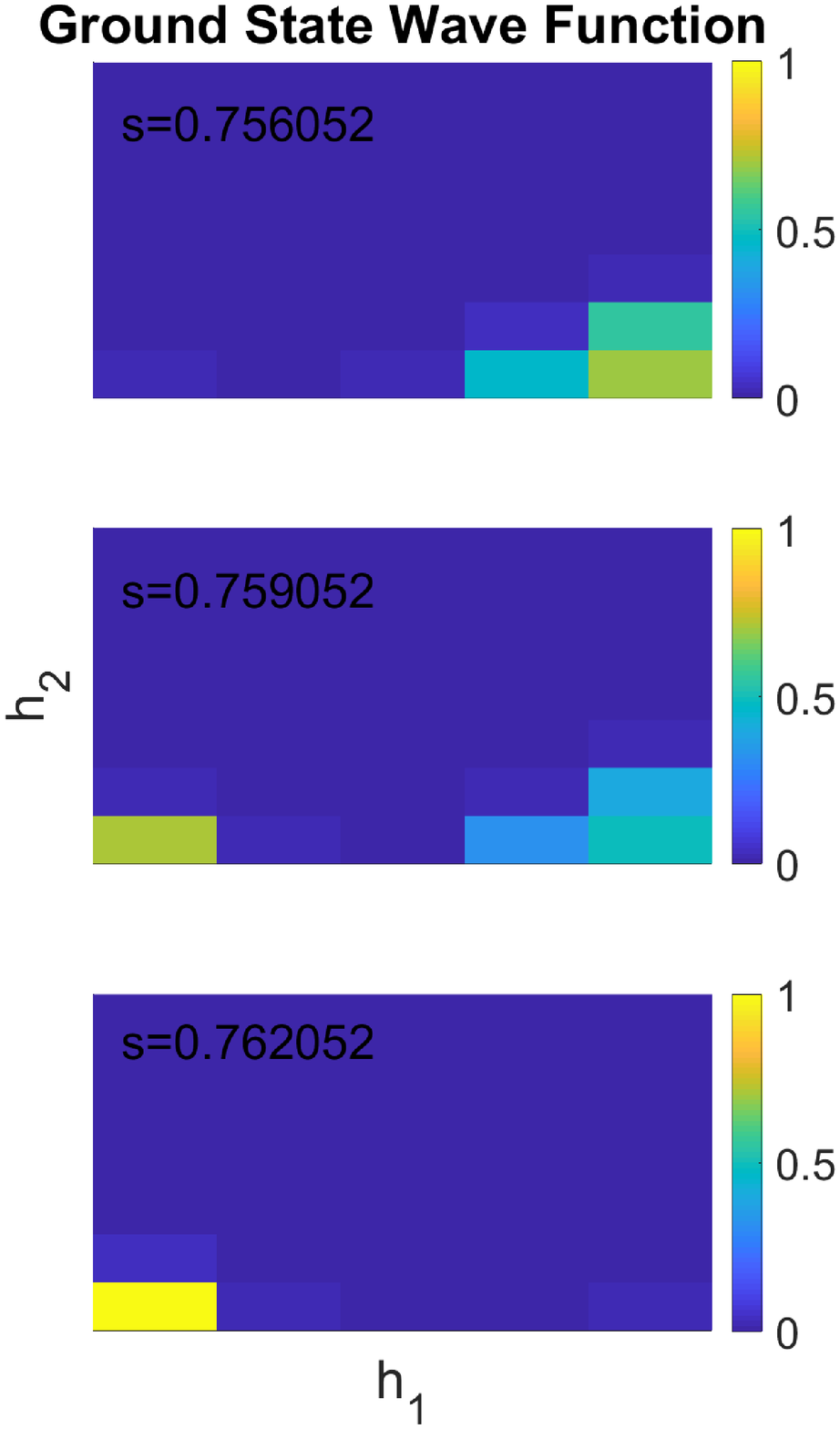}}
    \subfloat[]{ \includegraphics[width=0.5\columnwidth]{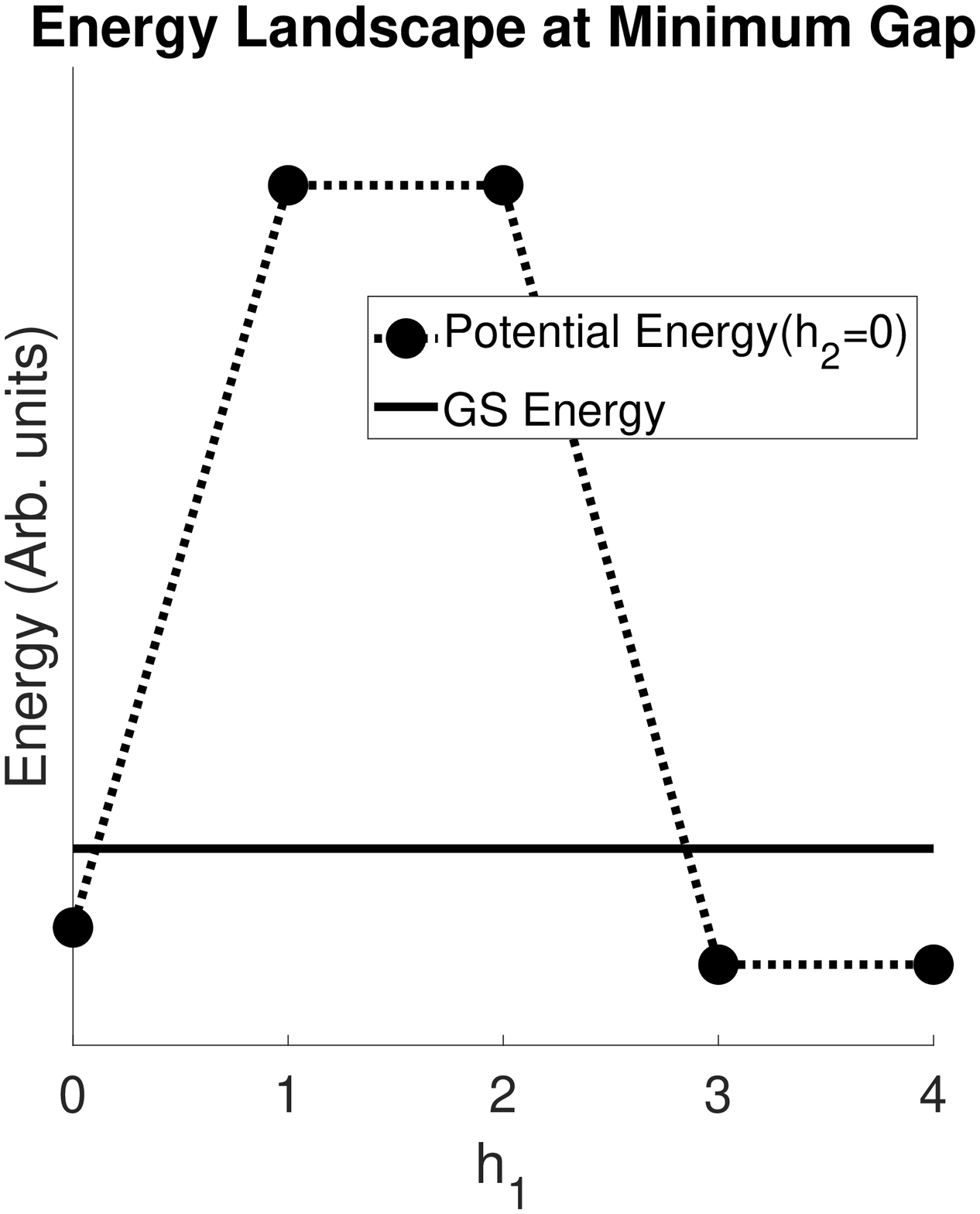}}
    \caption{(a) Exact ground state of a two well system on 10 qubits ($n_1=4$, $n_2=6$) as a function of $h_1$ and $h_2$ for $s$ near the minimum gap. The well on the left (right) has width 0 (1) and we see that at the point of minimum gap the wave function can rapidly switch between wells. (b) Energy landscape for fixed $h_2=0$ (cut with minimum distance between wells). The minimum eigenvalue gap occurs at an energy such that the parts of the wave functions external to the wells have amplitudes on bit strings for which the potential energy is greater than the eigenenergy of the state. Here potential energy is shifted by $-(1-s)I$ so that the relation to tunneling in the sense that the there is wave function support on bitstrings such that the energy is less than the potential energy is readily apparent.  }
    \label{fig:wf}
\end{figure}

\section{Numerical Considerations and Error Estimates}

Once we construct the tight-binding matrices $H^{(TB)}$ and $S$, we must solve the generalized eigenproblem given in $\ref{eqn:tbformula}$. This is complicated by the fact that $S$ may be ill-conditioned, leading to numerical instabilities. We address this by using the Fix-Heiberger reduction algorithm for solving symmetric ill-conditioned generalized eigenproblems \cite{fix1972algorithm}. In our code, we used a Lapack-style implementation of this algorithm from \cite{jiang2015fixnotes}. This algorithm works for real symmetric matrices with $S$ positive definite with respect to some user defined tolerance $0<\epsilon\ll 1$. Essentially this algorithm finds the eigenvalues of $H$ and $S$ which are zero with respect to $\epsilon$ and discards them before diagonalizing the remaining blocks of these matrices to solve the generalized eigenproblem.

 We expect the approximation to be good when the wave functions are ``tightly bound". Therefore we propose an approximate upper bound on the error based on first order pertubations of the diagonal elements of $H^{(TB)}$: $\Big[\sum_i |\bra{\psi_i}{\bar H^{(i)}}\ket{\psi_i}|^2\Big ]^{1/2}$ where $\bar H^{(i)}$ is the Hamiltonian excluding the potential from the $i^\mathrm{th}$ well. We test this upper bound on a set of 2450 random runs on between 4 and 10 qubits, between 2 and 10 wells, with depths between -1.0 and -5.99 (arbitrary units) and width 0. Our intuition is confirmed and we see that the relative error in the energy gap for zeroth order TB at 17 evenly spaced $s$ values is with high probability upper bounded by our proposed error estimate as shown in Fig. \ref{fig:randomDataRuns}. We estimate the relative error as 
 
 \begin{equation*}
\Big[\sum_i |\bra{\psi_i}\bar H^{(i)}\ket{\psi_i}|^2\Big ]^{1/2}/\tilde\gamma^{(TB)}     
 \end{equation*}
  where 
\begin{equation*}
  \tilde\gamma^{(TB)}:= E_1^{(TB)}-E_0^{(TB)}-\sqrt{2}\Big[\sum_i |\bra{\psi_i}{\bar H^{(i)}}\ket{\psi_i}|^2\Big ]^{1/2}
\end{equation*}
is the minimal possible gap within the absolute error estimates. Figure \ref{fig:empiricalCDF} shows the empirical CDF from the data, further indicating that this is with high probability a good upper bound. In particular problems, some of the points violating this upper bound can be eliminated as unstable eigenvalues using a case by case choice of $\epsilon$ in the Fix-Heiberger algorithm as for these runs $\epsilon$ was fixed at $0.1$ in all cases.

\begin{figure}[h]
    \centering
    \includegraphics[width=9cm,height=9cm,keepaspectratio]{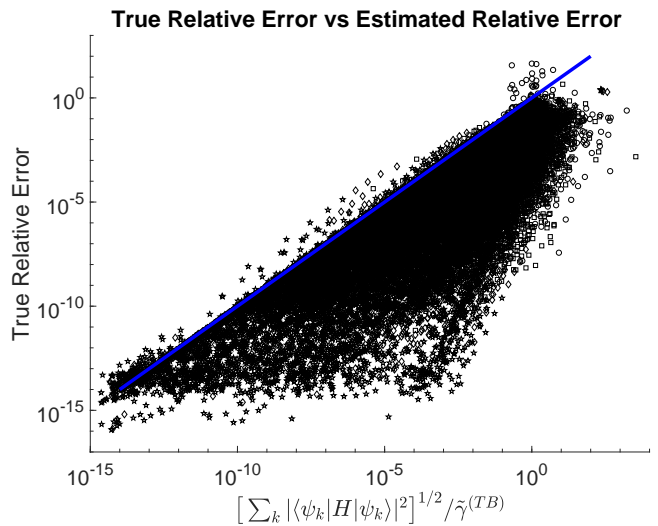}
    \caption{True error (tight-binding compared to direct diagonalization) versus error estimate for eigenvalue gap in zeroth order TB for a set of 2450 random runs on between 4 and 10 qubits, between 2 and 10 wells, with depths between -1.0 and -5.99 and width 0 at 17 evenly spaced $s$ values.  If $\tilde\gamma^{(TB)}<1$ then tight-binding cannot distinguish between the ground state and first excited state and the corresponding point is not plotted. Circles indicate $s\in[0.15, 0.30]$, squares $s\in[0.15, 0.30]$, diamonds $s\in[0.35, 0.70]$, and stars $s\in[0.70, 0.95]$. Note for the low $s$ points violating the upper bound estimate in the top right of the figure, with a different choice of $\epsilon$ the Fix-Heiberger algorithm could eliminate these points as unstable eigenvalues.}
    \label{fig:randomDataRuns}
\end{figure}

\begin{figure}[h]
    \centering
    \includegraphics[width=9cm,height=9cm,keepaspectratio]{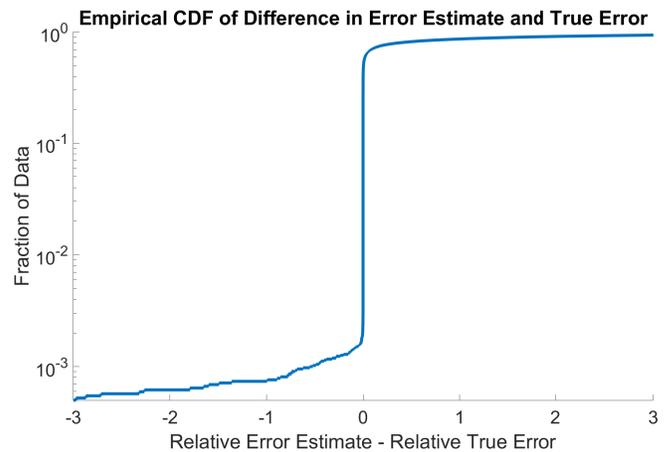}
    \caption{Error estimate minus true error estimated CDF for zeroth order TB for the 2450 random runs from Fig \ref{fig:randomDataRuns}.}
    \label{fig:empiricalCDF}
\end{figure}

Finally note that the scaling of the zeroth order tight-binding algorithm is $O(T(K^2n^6+K^3))$ where $T$ is the number of $s$ values for which the eigenspectrum is computed. The dominant factor comes from computing the $O(K^2)$ matrix elements of the tight-binding Hamiltonian, each of which requires diagonalizing a dimension $O(n^2)$ square matrix. Our codes for the exact solvers for $K\leq 3$ and for tight-binding along with input files for all the examples next presented are located on GitHub \cite{github}.
\end{section}

\begin{section}{Examples and Applications}
We now lay out a series of examples and possible applications aimed at revealing the power (and limits) of these algorithms. Example 1 gives the application of unstructured search with a prior guess, each represented as a single well. We solve this problem using both the exact two well algorithm and the tight-binding algorithm for up to a fairly large ($n=70$) number of qubits, demonstrating both of their effectiveness. Example 2 draws attention to how the tight-binding framework generates an effective graph (Hamiltonian) describing the problem, but it looks at this correspondence in reverse, by considering simulating an Ising model ground state adiabatically, using tight-binding as a tool for mapping the two systems to one another. The particular example presented has no practical implementation and serves merely as an interesting example of tight-binding with multiple wells, but we suggest generalizations beyond the scope of this paper that could prove useful. Example 3 demonstrates the effectiveness of tight-binding for a large number of wells ($n=10$, $K=50$). Finally, Example 4 highlights a situation where first order tight-binding is needed.

\subsection*{Example 1: Unstructured Search with Priors}
 Unsurprisingly, as AQC is equivalent to the standard circuit model of quantum computation with polynomial overhead \cite{Aharonov} an AQC version of Grover's algorithm for unstructured search demonstrates an equivalent speedup \cite{roland2002quantum}. The speedup requires having an optimized adiabatic schedule where the adiabatic condition is obeyed locally, speeding up when the eigenvalue gap between the ground state and first excited state is large and slowing down when the gap is small. If one ran the adiabatic algorithm purely at the rate prescribed by the minimum gap, the cost would be equivalent to classical unstructured search. Such an optimized schedule depends on having knowledge of the gap structure at $O(\log(1/\gamma))$ points throughout the evolution at a precision also $O(\log(1/\gamma))$ \cite{jarret2018quantum}, which is potentially a challenge for general problems. Questions have also been raised as to how robust this highly optimized evolution is to noise \cite{slutskii2019analog}, although static noise and time-dependent noise small relative to the gap can be handled \cite{jarret2018quantum}.
 
 Here we set aside these questions and the process of integrating over local gaps and simply investigate the minimum gap while searching for a single marked item as in Grover search but with a prior guess to the location of the marked item. If these estimates of the location of the marked item are good, then we expect the eigenvalue gap to be correspondingly wider, thus making it easier to find the marked item. This situation fits neatly into our model. We simply start at $s=0$ with a potential well representing our guess for the marked item and evolve to the potential giving the marked state. In principle, this ``guess" well could be tuned to have a functional form such that the initial wave function corresponds to a particular probability distribution corresponding precisely to our confidence in our initial guess. For simplicity, however, we shall treat our guess simply as a constant potential Hamming ball of some radius. This setup is given by the Hamiltonian

\begin{widetext}
\begin{equation}
    \label{eqn:groverwithpriors}
    H(s)=(1-s)\Big [-\frac{1}{n}\sum_j X_j + V_p \Theta \Big (r_p-\sum_j X^{\bar k}\bar Z_jX^{\bar k}\Big )\Big ]-s\ket{m}\bra{m}
\end{equation}
\end{widetext}
where $V_p<0$, $r_p$ is the depth and Hamming radius of the prior well $p$, respectively, and $\ket{m}$ is the marked item. 

With a single bit string prior plus the marked item we can use the exact two well solver and find exact solutions, so for our example we compare both this exact solution and the tight-binding solution. For simplicity, we consider point-like wells ($r_p=0$) so we only have to worry about the $\sigma=0$ subspace and zeroth order tight-binding. Figure \ref{fig:OneDeltaWell} shows how the eigenvalue gap scales with distance of one such prior from the marked item. Both the exact (two well subspace) and tight-binding solutions are shown, as further evidence of the accuracy of tight-binding for larger numbers of qubits. We see from these results that if the prior is close to the true marked item we do see an increase in the gap relative to standard Grover with no priors. However, if the prior is far from the true marked item, the gap shrinks relative to Grover with no prior.  

We should expect that randomly guessing will not provide any advantage (i.e. the prior must actually come from some prior knowledge about the problem). As shown in Figure \ref{fig:probabilityscaledgap} which plots the gap times the probability of randomly guessing a prior at the Hamming distance $R$ versus $n$ that with random guessing we recover a $O(2^{-n/2})$ scaling which we expect for unstructured search. 

\begin{figure}[h]
    \centering
    \includegraphics[width=8.5cm,keepaspectratio]{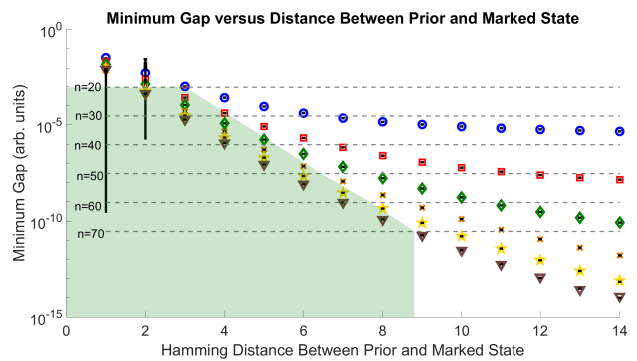}
    \caption{Plot of eigenvalue gap versus distance from marked item for a single width 0, depth -1 marked item for n=20 (circles), 30 (squares), 40 (diamonds), 50 (cross), 60 (five point star), 70 (triangle). Error bars overlaid on these points indicate the tight-binding results. Horizontal dotted lines show the gap for no prior and the shaded region indicates the distances where the prior offers an improvement in the gap over standard Grover.}
    \label{fig:OneDeltaWell}
\end{figure}

\begin{figure}[h]
    \centering
    \includegraphics[width=9cm,height=9cm,keepaspectratio]{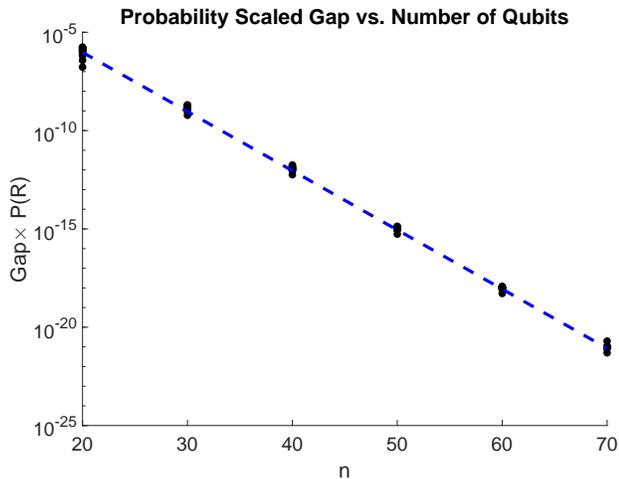}
    \caption{Plot of probability scaled minimum eigenvalue gap vs. number of qubits. The dashed line demonstrates that we recover $O(2^{-n/2})$ scaling if the priors are randomly guessed.}
    \label{fig:probabilityscaledgap}
\end{figure}

\begin{figure}[h]
    \centering
    \includegraphics[width=8.5cm,keepaspectratio]{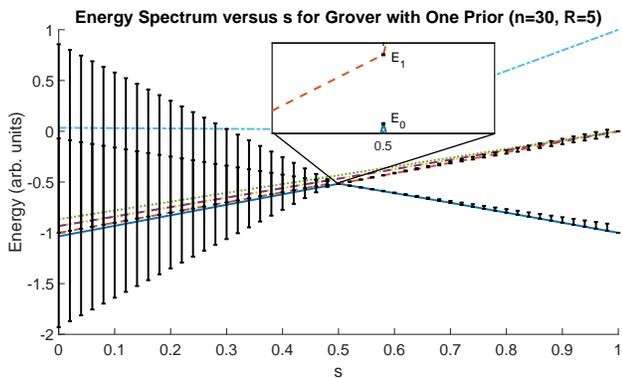}
    \caption{Plot of a particular eigenvalue spectrum vs. $s$ for the Grover with priors example. Lines indicate exact eigenvalues whereas error bars give the lowest two energy levels as determined by tight-binding. We see that tight-binding does an excellent job at finding the minimum gap in this problem with error bars on the order of the machine precision.}
    \label{fig:spectrumvss}
\end{figure}

In Figure \ref{fig:spectrumvss} we show a particular energy spectrum ($n=30$, $r_p=5$), demonstrating the size of the error bars in tight-binding as a function of $s$. We see that tight-binding does an excellent job of identifying the minimum gap with small error (at $s=0.5$). Also note from the exact solution that a number of other energy eigenvalues are all clustered around this point (many of which are highly degenerate). Such a feature could make these problems difficult for standard power iteration type methods for finding principal eigenvalues.

The next natural cases to consider are prior probability distributions favoring multiple bit strings and/or priors with $r_p>0$. However, we find that for this particular example tight-binding is a relatively poor approximation due to strong overlaps, as indicated by our error estimate. While these issues could be ameliorated by using deeper wells, to explore cases with multiple wells or wells with $r_p>0$ we turn to other examples.

\subsection*{Example 2: Approximate Ground State of a System of Strongly Interacting Spins}
A key feature of tight-binding is that it generates a greatly reduced effective graph (representing a Hamiltonian) with edge weights determined by the tunneling matrix elements between wells. One could imagine being given a real world Hamiltonian and then trying to identify these key features and approximately modeling it with tight-binding. This is a hard problem, however, so we leave this aside and work with a simpler example designed to demonstrate this key point. In particular, we work in reverse: starting with a small collection of strongly interacting spins we use the tight-binding framework to simulate the ground state of this Hamiltonian using the adiabatic Hamiltonian with potential wells considered in this paper. As an explicit case consider an Ising model on $L$ quantum spins of the form

\begin{equation}
    H_I=-\sum_{i<j}^LJ_{ij} Z_iZ_j - \sum_{i=1}^L B_i X_i - \alpha \mathbbm{1}
\end{equation}
where the last term is simply a potential shift chosen to make the diagonal terms strictly negative. If we can choose a collection of $2^L$ wells on some set of $n$ qubits with tight-binding Hamiltonian $H^{(\mathrm{TB})}(s^*)$ and overlap matrix $S(s^*)$ such that $[S^{-1}H^{(TB)}](s^*)\approx H_I$ then by evolving adiabatically from $s$ to $s^*$ we can approximately sample the ground state of $H_I$. 

Such a framework may have practical applicability, although understanding the extent of generality to a broad class of underlying Hamiltonians would require a thorough investigation of which interactions can be modeled using potential wells on a hypercube lattice. Namely we are limited to tunneling between wells, which enforces a fairly restrictive geometric dependence on the various matrix elements of the tight-binding Hamiltonian. There is still a number of available degrees of freedom, however, including the number of qubits in the host system, the shape and structure of the tight-binding wells, and global potential terms that could assist in tunneling between certain wells. Therefore, at least for certain problems, we conjecture that such a procedure could be practically useful in two cases: (1) if one has a large number of qubits that can be evolved adiabatically but a limited set of available controls (namely individual control in the computational basis but only a global $X$ term, as in the DWave machine \cite{dwave}), it would allow one to simulate more complicated interactions; (2) depending on how robust the tight-binding framework is to noise in the underlying system of qubits this scheme could serve as a method for fault-tolerant simulation on a large number of noisy qubits. The Davis-Kahan theorem suggests that our framework is indeed robust to moderate noise \cite{davis1970rotation}. We note that the Hamming ball wells used here are not easily generated on current hardware as they require $n$-body interactions. However, such potential wells are a simplification for ease of testing our code. One could make $K=2^L$ wells with an $K$-local potential using a degree $2^K$ polynomial in the distances to each of the desired wells.

Here we leave these more general questions open and raise them purely as a possible motivation for a particular example which demonstrates the effectiveness of tight-binding. For simplicity assume that $B_i\ll J_{ij} \, \forall \, i,j$, which means that $S\approx\mathbbm{1}$ and our mapping is simply $H^{(\mathrm{TB})}(s^*)=H_I$ for some $s^*$. Then consider a system of 3 spins with interactions $J_{ij}=1 \,\forall\, i,j$ and $B_i=0.015 \,\forall\, i$ (arbitrary units). Using $\alpha=30$ and $s^*=0.95$, this Hamiltonian can be mapped to a tight-binding Hamiltonian of 8 hyperspherical wells (2 with $s^*V_p=-33.5$ and 6 with $s^*V_p=-29.5$, $r=0$) on 10 qubits where we consider terms in the tight-binding Hamiltonian smaller than $O(10^{-5})$ as essentially zero. In Figure \ref{fig:probabilitycomparison}, we compare the exact (full diagonalization) adiabatic ground state probability distribution at $s^*$ as determined via the tight-binding mapping and the exact Ising model ground state probability distribution. The adiabatic probability is determined by the normalized probability of sampling within the well corresponding to a given Ising model basis state. We see that the mapping is a good one: the adiabatic version samples with probabilities within $O(10^{-5})$ of the true probability. 

\begin{figure}[h]
    \centering
    \includegraphics[width=9cm,height=9cm,keepaspectratio]{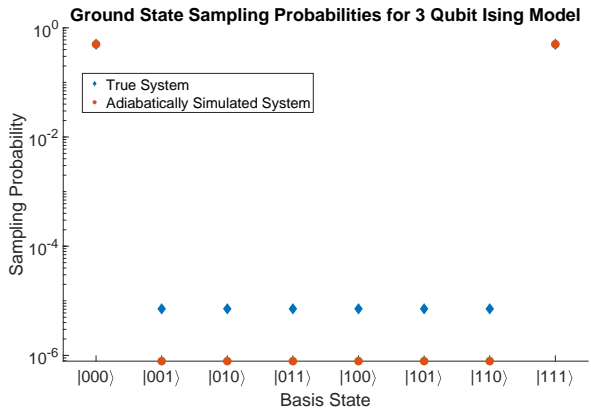}
    \caption{Comparison of exact adiabatically simulated ground state probability distribution to exact true ground state for the 3 qubit Ising model with $J_{ij}=1 \,\forall\, i,j$ and $B_i=0.015 \,\forall\, i$. The adiabatic Hamiltonian is constructed via a mapping between tight-binding with 8 wells and the 3 qubit Ising model Hamiltonian.}
    \label{fig:probabilitycomparison}
\end{figure}

\subsection*{Example 3: Large Set of Wells}
As a final test for a large set of wells, we compared exact diagonalization and tight-binding for a set of 50 wells with depths between $-1$ and $-1.13$ with $r=0$ on 10 qubits, as depicted in Fig. \ref{fig:example3}. We found that tight-binding was effective in this case, however, the point in the evolution in $s$ where the wave function is tightly bound enough for the errors to be small enough to identify the gap is later than in problems with fewer wells. This suggests that for large sets of wells tight-binding is only useful at identifying the minimum gap when the minimum gap occurs late in the evolution, as in the example here, where the gap is at $s=1$. This brings attention to one key limitation of tight-binding: it does not let us know if the minimum gap is prior to the point in the schedule where the tight-binding errors decrease to the point that we can resolve the gap. Figure \ref{fig:example3} also depicts the results for tight-binding without the use of the Fix-Heiberger algorithm. We see that due to numerical instability the eigenvalues diverge and Fix-Heiberger allows us to be control this divergence and to automatically eliminate unstable eigenvalues with appropriate choice of $\epsilon$.

\begin{figure}[h]
    \centering
    \includegraphics[width=9cm,height=9cm,keepaspectratio]{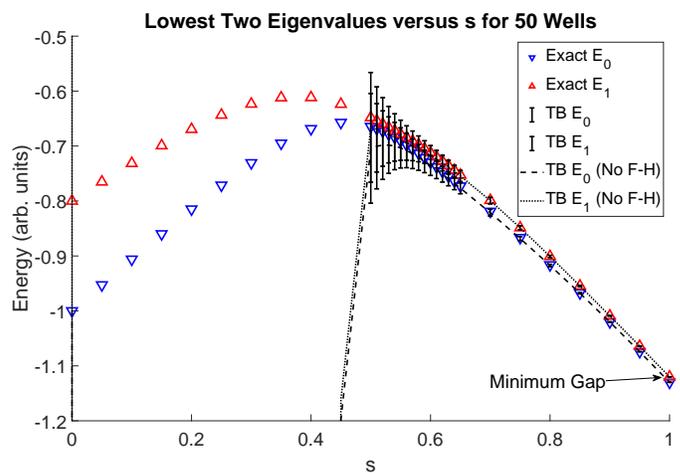}
    \caption{Plot of lowest two eigenvalues versus s for Example 3. We see at large $s$ tight-binding is effective, however, a large number of wells does require that the minimum gap occur at larger $s$ to capture it effectively with tight-binding. Lines indicate tight-binding without Fix-Heiberger indicating its usefulness to automatically remove unstable eigenvalues.}
    \label{fig:example3}
\end{figure}

\subsection*{Example 4: First Order Tight-binding}
Problems involving wells with $r>0$ possibly requires first order tight-binding in order to capture internal structure of the wider wells and to check whether or not these states are relevant. Such problems are not quite as easily automated as zeroth order tight-binding and tend to require more fine tuning of the $\epsilon$ parameter in the Fix-Heiberger reduction algorithm, as well as comparison between zeroth and first order solutions to correctly ascertain the appropriate gap. Here we consider an example designed to demonstrate the need for first order tight-binding. In particular, we consider two Hamming spherical potential wells 

\begin{equation}
    H(s)=-\frac{1-s}{n}\sum_j X_j -s\sum_{i\in\{a,b\}}V_i\Theta \Big (r_i-\sum_j X^{\bar i}\bar Z_jX^{\bar i}\Big )
\end{equation}

with $V_a=-5$, $V_b=-4.9$, $r_a=1$, $r_b=0$, $|\bar a|-|\bar b|=6$, and $n=10$. In this case, due to well $a$ being energetically favored for all $s$ the full first excited state wave function is constructed using the first excited state of well $a$. Zeroth order tight-binding doesn't capture this behavior. This is shown in Table \ref{tbl:schedule2ndOrderExample}. Note that in this calculation we did not use the first excited state of the width 0 well, as this state is unnecessary and not tightly bound and thus introduces avoidable error into the approximation. 

In most problems with Hamming symmetric wells, this is not an issue and even with wider wells one can get the correct first excited state purely with zeroth order tight-binding, but it is important to have these tools available to have a complete understanding of the energy spectrum.

\begin{table}[]
\begin{tabular}{cccc}
s    & $E_1$ (Exact) & $E_1$ (TB0) & $E_1$ (TB1) \\
0.20    & -1.05367      & -1.04988    & -1.05350    \\
0.30   & -1.52649      & -1.50398    & -1.52649    \\
0.40    & -2.01448      & -1.73993    & -2.01448    \\
0.50   & -2.50802      & -2.46024    & -2.50802    \\
0.60     & -3.00427      & -2.94545    & -3.00427    \\
0.70    & -3.50206      & -3.43262    & -3.50206    \\
0.80     & -4.00080      & -3.92102    & -4.00080    \\
0.90    & -4.50018      & -4.41022    & -4.50018    \\
\end{tabular}
\caption{\label{tbl:schedule2ndOrderExample}First excited state energies for Example 4. First order tight-binding is needed to get the correct solution in this problem.}
\end{table}

\end{section}

\begin{section}{Conclusion}
We provide a set of algorithms for efficiently analyzing the performance of AQC for problems that can be expressed in terms of a set of individually Hamming symmetric wells. These problems, while still highly symmetric, are more complex than the well studied Hamming symmetric example and should provide a new testbed for study of AQC. In particular, the tight-binding approach for studying this model highlights the effects of tunneling, which must be the source of the quantum speedup if one is afforded by AQC with stoquastic Hamiltonians We also provide several examples demonstrating the effectiveness of tight-binding as a tool for studying this toy model.

\end{section}

\section*{Acknowledgements} J.B. was supported by the DOE CSGF program (award No. DE-SC0019323) and also partially supported by NSF PFCQC program, DoE ASCR FAR-QC (award No. DE-SC0020312), DoE BES QIS program (award No. DE-SC0019449), DoE ASCR Quantum Testbed Pathfinder program (award No. DE-SC0019040), NSF PFC at JQI, ARO MURI, ARL CDQI, and AFOSR.

\bibliography{bib}

\appendix
\section{Fully-symmetric, stoquastic Hamiltonians}
\label{app:HammingSymmetric}
Consider a Hamming-symmetric, stoquastic Hamiltonian on $n$ qubits of the form
\begin{equation}
\label{eqn:hammingsymmetricHamiltonian}
H(s)=-\frac{a(s)}{n}\sum_j X_j+b(s)V\Big (\sum_j \bar Z_j \Big ),
\end{equation}
where $a(s)$ and $b(s)$ define the adiabatic schedule as a function of parameter $s\in[0,1]$ with the constraint that $a(0)=b(1)=1$ and $a(1)=b(0)=0$, and $\bar Z=(I-Z)/2=\ket 1\bra 1$ is the Hamming weight operator. Since the potential $V$ is a function only of Hamming weight, the Hamiltonian can be written in a basis such that it is block diagonal \cite{kechedzhi2016open, isakov2016understanding}. 

A Hamiltonian of this form commutes with the operator $J^2=J_x^2+J_y^2+J_z^2$ where
\begin{align}
J_x&=\frac{1}{2}\sum_j X_j & J_y&=\frac{1}{2}\sum_j Y_j & J_z&=\frac{1}{2}\sum_j Z_j,
\end{align}
so $J^2$ is a conserved quantity and we can use a basis of states $\ket{j m \gamma}: J^2\ket{j m \gamma}=j(j+1)\ket{j m \gamma}$ with total spin $j$ and $z$-projections $m\in\{-j, -j+1, ...,j-1, j\}$ with $\gamma$ labeling the degeneracies. Introducing a parameter $\sigma\in\{0,1,...,\floor{\frac{n}{2}}\}$ to label each total spin subspace, we have $j=\frac{n}{2}-\sigma$. The degeneracy for a given total spin $j$ and $z$-projection $m$ is determined by the number of representations of the group of permutations: $\Gamma=\binom{n}{\sigma}-\binom{n}{\sigma-1}$. Noting that $m$ can be written in terms of Hamming weight $w$ as $m=w-\frac{n}{2}$ we can express the basis states as
\begin{equation}
    \ket{w \sigma \gamma}=\frac{1}{\sqrt{\binom{n}{w}}}\sum_{|x|=w}k(x)\ket{x}.
\end{equation}
Here, the quantities $k(x)$ are numerical constants defining the weights of the appropriate bit strings so that we have an orthonormal basis. For the $\sigma=0$ ($j=n/2$) subspace $k=1$ $\forall$ $x$.

Upon defining raising and lowering operators $J^{\pm}=J_x\pm i J_y$ and noting $\sum_j X_j=J^++J^-$, one finds
\begin{eqnarray}
H&=&\sum_{w,\sigma,\gamma}\bigg [ -\frac{a(s)}{n}\Big [C^+\ket{w, \sigma, \gamma}\bra{w+1, \sigma, \gamma}\nonumber \\
&{ }&+C^-\ket{w\, \sigma\, \gamma}\bra{w-1\, \sigma\, \gamma}\Big ]\nonumber \\
&{ }&+b(s)V(w)\ket{w\, \sigma\, \gamma}\bra{w\, \sigma\, \gamma} \bigg ],
\end{eqnarray}
where 
$C^+=\sqrt{(w-\sigma+1)(n-\sigma-w)}$ and $C^-=\sqrt{(w-\sigma)(n-\sigma-w+1)}$ are the standard raising and lowering coefficients. For a given subspace $\sigma\in\{0,1,...,\floor{\frac{n}{2}}\}$, $w\in[\sigma, n-\sigma]$. This Hamiltonian is block diagonal with each block corresponding to a subspace. The $\sigma=0$ block is a permutation-symmetric block of dimension $(n+1)\times(n+1)$. As permutation of bits is a symmetry of the Hamiltonian in \ref{eqn:hammingsymmetricHamiltonian}, and by the Perron-Frobenius theorem the ground state is non-degenerate, the ground state of the Hamiltonian must exist in this block. Therefore the only relevant gap is within this subspace, so one can directly diagonalize this permutation symmetric subspace to analyze the performance of AQC on Hamiltonians of the form of Eq.~(\ref{eqn:hammingsymmetricHamiltonian}) up to a large number of qubits. 

\section{Relabeling bases for three or fewer Hamming-symmetric wells}
\label{app:few}
Consider Hamiltonians such as given by Eq.~(\ref{eqn:generalHamiltonian}),
\begin{equation*}\label{app:eqn:generalHamiltonian}
    H(s)=-\frac{a(s)}{n}\sum_j X_j + \sum\limits_{k=0}^K b_k(s)V_k\Big (\sum_jX^{\bar k}\bar Z_jX^{\bar k}\Big ),
\end{equation*}
for the simplest non-trivial cases, $K=2,3$. For these two cases, it is possible to relabel the bit strings in a manner which preserves Hamming separations and makes possible efficient exact calculations of the relevant spectral gaps.

${\bf K=2:}$ Given any two $n$-bit strings $\beta_a$ and $\beta_b$ we can always introduce a relabeling that preserves the Hamming distance $n_1$ between them, such that these two bit strings have the form
\begin{eqnarray}
\label{app:2wells}
\beta_a & = &\ket{0..0\;\;0..0} \nonumber \\
\beta_b & = &|\underbrace{1..1}_{n_1}\underbrace{0..0}_{n_2}\rangle.
\end{eqnarray} 
By construction, $n_1$ is the Hamming distance between the bit strings and $n_2=n-n_1$. We identify the subset of the first $n_1$ relabeled bits as $S_1$ and the remaining $n_2$ bits as the subset $S_2=\bar S_1$. For a general relabeled bit string,  define $h_1$ to be the number of ones in $S_1$, and the quantity $h_2$ the number of ones in $S_2$. For each subset, the parameters $\sigma_1, \sigma_2, \gamma_1,$ and $\gamma_2$ may be defined as in the Hamming symmetric case. This defines a new labeling of basis states given by $\ket{h_1 h_2 \sigma_1 \sigma_2 \gamma_1 \gamma_2}$.

From the Perron-Frobenius theorem and symmetry we know the ground state exists in the $\sigma_1=\sigma_2=0$ subspace. In particular, the Perron-Frobenius theorem guarantees a non-degenerate ground state with non-negative amplitudes. Additionally, the symmetry group for this Hamiltonian is the direct sum of symmetric groups $S_{n_1}\oplus S_{n_2}$ acting on the sets $S_1$ and $S_2$ respectively. The trivial representation is associated with our product state basis $\ket{h_1 h_2 \sigma_1 \sigma_2 \gamma_1 \gamma_2}$ for $\sigma_1=\sigma_2=\gamma_1=\gamma_2=0$. However we could also write a basis diagonal in the total ``spin" $\sigma$ which for $\sigma=0$ is the 1D representation of the group consistent with the non-negative amplitude requirements of the Perron-Frobenius theorem. Therefore the ground state is guaranteed to transform within this one-dimensional representation group of the Hamiltonian. As $\sigma_1+\sigma_2\leq\sigma$ this subspace is fully within the $\sigma_1=\sigma_2=0$ subspace in the product basis.  

Therefore, as the ground state exists in this $\sigma_1=\sigma_2=0$ subspace, the relevant gap is also in this subspace, which has dimension $(n_1+1)(n_2+1)\times (n_1+1)(n_2+1)$. The relevant gap can be exactly computed by direct diagonalization of a matrix of dimension polynomial in both $K$ and $n$.

With $\sigma$ and $\gamma$ labels dropped for compactness the Hamiltonian in this space can be exactly written as 

\begin{align}
\label{eqn:twowellH}
H=&\sum_{h_1 h_2 \sigma_1 \sigma_2 \gamma_1 \gamma_2}\bigg [-\frac{a(s)}{n}\Big [C^+_1\ket{h_1 h_2}\bra{h_1+1\, h_2} \nonumber\\
&+C^-_1\ket{h_1 h_2}\bra{h_1-1\, h_2}+C^+_2\ket{h_1 h_2}\bra{h_1\, h_2+1}\nonumber\\
&+C^-_2\ket{h_1 h_2}\bra{h_1\, h_2-1}\Big ] +\sum_{i\in\{1,2\}}b_i(s)V_i(h_1, h_2)\bigg ]
\end{align}
where $C^\pm$ act just on the relevant subset.

${\bf K=3:}$ For this case, the three selected bit strings are relabeled as follows:
\begin{align}
\label{app:3wells}
\beta_a & = &\ket{1..1\ \, 0..0\ \, 0..0\ \, 0..0} \nonumber\\
\beta_b & = &\ket{0..0\ \, 1..1\ \, 0..0\ \, 0..0} \nonumber\\
\beta_c & = &|\underbrace{0..0}_{n_1}\underbrace{0..0}_{n_2}\underbrace{1..1}_{n_3}\underbrace{0..0}_{n_4}\rangle. 
\end{align}
As was the case for two wells, a basis of the form $\bigotimes_{ k=\{1,2,3,4\} }\ket{h_k \sigma_k \gamma_k}$ exists. Again via symmetry of the Hamiltonian, the ground state exists in the $\sigma_k=0\, \forall\, k$ subspace, whose dimension is polynomial in $K$ and $n$.

\section{Tight-binding matrix elements}
\label{app:tbbasis}
Consider \ref{eqn:htb} reproduced here
\begin{align}
    &H^{(TB)}_{ij}= \\ 
    &\bra{\psi_i}H_d + \sum_{h_1 h_2}[ V_i(h_1, h_2) + V_j(h_1, h_2) + V_c(h_1,h_2)]\ket{\psi_j} \nonumber
\end{align}
where $H_d$ is the driver part of the Hamiltonian in the appropriate basis $V_i$ and $V_j$ are the diagonal potential terms corresponding to the $i^{\mathrm{th}}$ and $j^{\mathrm{th}}$ wells and

\begin{equation}
V_c=\frac{\sum_{k\neq i,j}\sum_{r_k=0}^{n} N(h_1, h_2,n_1,R_{ik},R_{jk},r_k)V_k(r_k)}{\sqrt{\binom{n_1}{h_1}\binom{n-n_2}{h_2}}}.
\end{equation}

Note that this is equivalent to \ref{eqn:twowellH} with the additional correction factor $V_c$. Recall that $n_1$, $R_{ik}$ and $R_{jk}$ are the Hamming distance between the $i^{th}$ and $j^{th}$ wells, the $i^{th}$ and $k^{th}$ wells and the $j^{th}$ and $k^{th}$ wells, respectively, $r_k$ is the distance from the $k^{th}$ well and $V_k(r_k)$ is the potential due to the $k^{th}$ well at distance $r_k$. The function $N(h_1, h_2,n_1,R_{ik},R_{jk}, r_k)$ gives the number of points of intersection between Hamming spheres of radius $r_i=h_1+h_2$ and $r_j=h_2+(n_1-h_1)$ centered on the $i^{th}$ and $j^{th}$ wells respectively and the Hamming sphere of radius $r_k$ centered on the $k^{th}$ well.

To find the function $N(h_1, h_2,n_1,R_{ik},R_{jk},r_k)$, without loss of generality consider 3 wells $i,j,k$ shifted so they are in the form of \ref{app:3wells} where we label the corresponding sets of qubits as $n_1', n_2', n_3', n_4'$ to differentiate from $n_1$ and $n_2$ in the 2 well basis for wells $i$ and $j$. 

\begin{align}
n_1=n_1'+n_2' && R_{ik}=n_1'+n_3' && R_{jk}=n_2'+n_3'.
\end{align}
which we can solve for $n_1', n_2', n_3'$ in terms of the input parameters.

Define $h_1', h_2', h_3', h_4'$ as the number of ones in each of the 4 subsets of qubits. Therefore

\begin{align}
r_i&=(n_1'-h_1')+h_2'+h_3'+h_4' \nonumber \\ r_j&=h_1'+(n_2'-h_2')+h_3'+h_4' \nonumber \\ r_k &= h_1'+h_2'+(n_3'-h_3')+h_4' .
\end{align}

This is a system of linear Diophantine equations whose solutions satisfy

\begin{align}
h_2'&=\frac{1}{2}(n_2'+r_i-r_j-n_1')+h_1'\\
h_3'&=\frac{1}{2}(n_3'+r_i-r_k-n_1')+h_1'\\
h_4'&=\frac{1}{2}(n_2'+n_3'-r_j-r_k)-h_1'.
\end{align}

It is straightforward to then count the solutions for $h_1'\in [0, n_1']$. We try each possible $h_1'$ and check that $h_2'\in[0,n_2']$ and $h_3'\in[0,n_3']$. Each solution found is then multiplied by the combinatoric factor $\binom{n_1'}{h_1'}\binom{n_2'}{h_2'}\binom{n_3'}{h_3'}\binom{n_4'}{h_4'}$. The total result is $N$. 

\section{Proof of Possible Subspaces for First Excited State}
\label{app:subspacepf}
Here we prove that the first excited state for a single Hamming symmetric well must exist in either the $\sigma=0$ or $\sigma=1$ subspace. Consider an eigenstate $\ket\psi=\sum_{w\sigma\gamma}\alpha(w, \sigma, \gamma)\ket{w\sigma \gamma}$. Then

\begin{widetext}
\begin{align}
    H\ket{\psi}=\sum_{w\sigma\gamma}\Bigg [-\frac{1-s}{n}\bigg (\alpha(w+1, \sigma, \gamma)\sqrt{(w-\sigma+1)(n-\sigma-w)}+\alpha(w-1, \sigma, \gamma)&\sqrt{(w-\sigma)(n-\sigma-w+1)} \bigg )\nonumber\\
    &-s\alpha(w, \sigma, \gamma)V(w)\Bigg ]\ket{w\sigma \gamma}
\end{align}
\end{widetext}

which implies that for $\alpha(w, \sigma, \gamma)\neq 0$ that 
\begin{equation}
    E(s)=-\frac{1-s}{n}\big (r^+C^+ + r^- C^- \big )+sV(w)
\end{equation}
where $r^{\pm}=\frac{\alpha(w\pm 1, \sigma, \gamma)}{\alpha(w, \sigma, \gamma)}$ for all $w, \sigma, \gamma$ and $C^{\pm}$ are the raising and lowering coefficients. Now consider the energy difference between candidate first excited states with different $\sigma$. The potential term is independent of $\sigma$ and thus does not affect which subspace is energetically favored. For a subspace $\sigma$ and a subspace $\sigma'>\sigma$,  $C^{\pm}(\sigma'>\sigma)<C^{\pm}(\sigma)$ so if $r^{\pm}$ were independent of $\sigma$ then the $\sigma=0$ subspace would always be favored. Now consider the difference in energy between the candidate first excited states:

\begin{align}
    \Delta E(s)&=E_{\sigma'}(s)-E_{\sigma}(s)\nonumber\\
    &=\frac{1-s}{n}[r^+_\sigma C^+_\sigma+r^-_\sigma C^-_\sigma-r^+_{\sigma'} C^+_{\sigma'}-r^-_{\sigma'} C^-_{\sigma'}]
    \label{deltaE}
\end{align}

The above equation is true for all $w$. We take $w=\sigma'$ so $C^-_{\sigma'}(w=\sigma')=0$ eliminating one term. For $\sigma >0$, $r^\pm$ must be nonnegative (by the Perron-Frobenius theorem) so $\Delta E(s)$ is nonnegative unless $r^+_{\sigma'}$ is large relative to $r^\pm_{\sigma}$. We will now show a contradiction. Consider the $w=\sigma'$ element of the eigenvector equation in both subspaces. In the $\sigma'$ subspace

\begin{align}
    -&\frac{1-s}{n}C^+_{\sigma'}(w=\sigma')\alpha(w=\sigma'+1, \sigma')\nonumber \\
    &+sV(w=\sigma')\alpha(w=\sigma', \sigma')=E_{\sigma'}\alpha(w=\sigma', \sigma')
\end{align}
and in the $\sigma$ subspace
\begin{align}
    -\frac{1-s}{n}&\Big [ C^+_{\sigma}(w=\sigma)\alpha(w=\sigma'+1, \sigma)+C^-_{\sigma}(w=\sigma)\nonumber\\ 
    &\times \alpha(w=\sigma'-1, \sigma) \Big ]+sV(w=\sigma')\alpha(w=\sigma', \sigma)\nonumber\\
    &=E_{\sigma}\alpha(w=\sigma', \sigma)
\end{align}
Rearranging and dropping the arguments of the functions for compactness we obtain from the fact $C^+_{\sigma'}<C^+_\sigma$
\begin{align}
    \big ( E_{\sigma'}-sV \big )\frac{1}{r^+_{\sigma'}}> \big ( E_{\sigma}-sV \big )\frac{1}{r^+_{\sigma}}+\frac{1-s}{n}C^-_\sigma\frac{r^-_\sigma}{r^+_\sigma}.
\end{align}
The last term is positive definite so we can drop it and get the inequality
\begin{equation}
    \frac{E_{\sigma'}-sV }{E_{\sigma}-sV }>\frac{r^+_{\sigma'}}{r^+_\sigma}.
\end{equation}
Both sides are positive definite for $\sigma,\sigma'>0$. And as $sV$ is the same in both subspaces, if $E_{\sigma'}<E_\sigma$ then $r^+_{\sigma'}<r^+_\sigma$ but this contradicts the result from \ref{deltaE} that for this to be true  $r^+_{\sigma}<r^+_{\sigma'}$. Therefore the first excited state must always exist either in the $\sigma=0$ or $\sigma=1$ subspaces. 

\end{document}